\documentclass[12pt,fleqn]{article}

\usepackage{amssymb,graphicx,epsfig}
\usepackage{lscape,graphics,amsmath,geometry}
\usepackage{latexsym}
\usepackage[dvips]{color}
\newcommand {\beq} {\begin{eqnarray}}
\newcommand {\eeq} {\end{eqnarray}}
\def\lsim{\raise0.3ex\hbox{$\;<$\kern-0.75em\raise-1.1ex\hbox{$\sim\;$}}}
\def\gsim{\raise0.3ex\hbox{$\;>$\kern-0.75em\raise-1.1ex\hbox{$\sim\;$}}}
\def\greaterthansquiggle{\raise.3ex\hbox{$>$\kern-.75em\lower1ex\hbox{$\sim$}}}
\def\lessthansquiggle{\raise.3ex\hbox{$<$\kern-.75em\lower1ex\hbox{$\sim$}}}

\newcommand{\ci}{\cite}
\newcommand{\beqn}{\begin{eqnarray}}
\newcommand{\eeqn}{\end{eqnarray}}
\newcommand{\bequ}{\begin{equation}}
\newcommand{\eequ}{\end{equation}}
\newcommand{\bsl}{\begin{sloppypar}}
\newcommand{\esl}{\end{sloppypar}}

\begin{document} 
\null
\hfill hep-ph/0211039\\
\vskip .4cm

\begin{center}
{\Large\bf
Disentangling fundamental MSSM Parameters:\\[.3em] 
Light Gaugino/Higgsino System\footnote[1]{Talk 
given at SUSY02, Hamburg, June 17-23, 2002}}
\vskip 1.5em

{\large
{
Gudrid Moortgat-Pick\footnote[7]{gudrid@mail.desy.de}
}                     
}\\[3ex]

{\footnotesize \it 
\noindent
DESY, Deutsches Elektronen-Synchrotron, D-22603 Hamburg, Germany,\\
II. Institut f\"ur Theoret. Physik, Universit\"at Hamburg, D-22761 Hamburg, 
Germany 
}\\
\end{center}
\vskip .5em
\par

\begin{abstract}
In order to reveal the underlying structure of Supersymmetry one
has to determine the low--energy parameters without assuming a
specific SUSY breaking scheme. In this paper we show a procedure how
to determine $M_1$, $\Phi_{M_1}$, $M_2$, $\mu$, $\Phi_{\mu}$ and 
$\tan\beta$ even in the case
when only light charginos $\tilde{\chi}^{\pm}_1$ 
and neutralinos $\tilde{\chi}^0_1$, $\tilde{\chi}^0_2$ would be 
accessible at the first stage of a future 
Linear Collider with polarized beams.
\end{abstract}

\section{Introduction}
\vspace{-.3cm}
Supersymmetry is one of the best motivated extensions of the Standard
Model (SM). Since, however, SUSY has to be broken the unconstrained
version of the Minimal Supersymmetric extension of the Standard Model
(MSSM) leads to about 105 new parameters to express all
possible soft breaking terms. In order to reveal the breaking
mechanism one has to determine these fundamental parameters in future
experiments without assuming a specific structure of the breaking
mechanism.

We concentrate in this paper on the gaugino/higgsino system, charginos
and neutralinos, which are the SUSY partners of the neutral and
charged vector and Higgs bosons. The mixing between these particles
depends on the fundamental U(1), SU(2) and higg\-sino mass parameters
including CP--violating phases -- $M_1$, $\Phi_{M_1}$, $M_2$, $\mu$,
$\Phi_{\mu}$ -- and the ratio of the two Higgs vacuum expectation
values $\tan\beta=v_2/v_1$. Several strategies have already been
worked out for determining these parameters at a future Linear
Collider in the energy range of $\sqrt{s}=500$--1000~GeV, as
e.g. TESLA \ci{TDR}. We demonstrate in this paper a procedure for
determining the parameters in the case that only the light states
$\tilde{\chi}^{\pm}_1$, $\tilde{\chi}^0_{1,2}$ would be accessible at the
first stage of a LC (see \cite{CKMZ}).

\section{MSSM Parameter Determination}
\vspace{-.3cm}
\subsection{Chargino Sector}
\vspace{-.1cm}
The $2\times 2$ chargino mixing depends on the parameters $M_2$, $\mu$,
$\Phi_{\mu}$ and $\tan\beta$. It can be described by two mixing
angles. The mixing angles can be determined e.g. when studying
cross sections with longitudinally polarized beams 
\cite{Choi}. Inversion of
the equations leads to the determination of the 
parameters with a sign ambiguity in $\Phi_{\mu}$, 
even if both $m_{\tilde{\chi}^{\pm}_{1,2}}$ are known. 
\subsection{Neutralino Sector}
\begin{figure}
\setlength{\unitlength}{1cm}
\begin{center}
\begin{picture}(15,6)
\put(0,0){\includegraphics{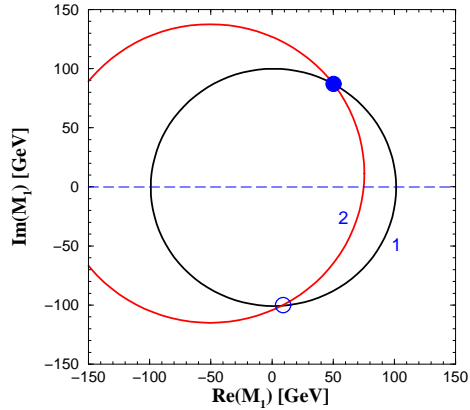}}
\end{picture}
\end{center}\vspace*{-2.6cm}
  \caption{\label{fig_1} Contours in the $Re(M_1)-Im(M_1)$ plane for two
  measured masses $m_{\tilde{\chi}^0_{1,2}}$. The other 
MSSM parameters are $M_2=190.8$~GeV, $|\mu|=365.1$~GeV, 
$\Phi_{\mu}=\pi/8$, $\tan\beta=10$ \ci{CKMZ,SPS}.}%
\end{figure}

The neutralino mixing depends -- in addition to the
parameters of the chargino system -- on the parameters $M_1$,
$\Phi_{M_1}$. The characteristic equation of the mass matrix squared,
$M M^{\dagger}$, can be written as a second order polynomial in
$M_1$. Therefore we are left with a two--fold ambiguity for $M_1$,
$\Phi_{M_1}$, when only exploring the two lightest masses
$\tilde{\chi}^0_1$, $\tilde{\chi}^0_2$, Fig.~\ref{fig_1} \cite{CKMZ}.

\begin{figure}[t]
\setlength{\unitlength}{1cm}
\begin{minipage}{7cm}
\begin{center}
\begin{picture}(15,6)
\put(0,0){\includegraphics{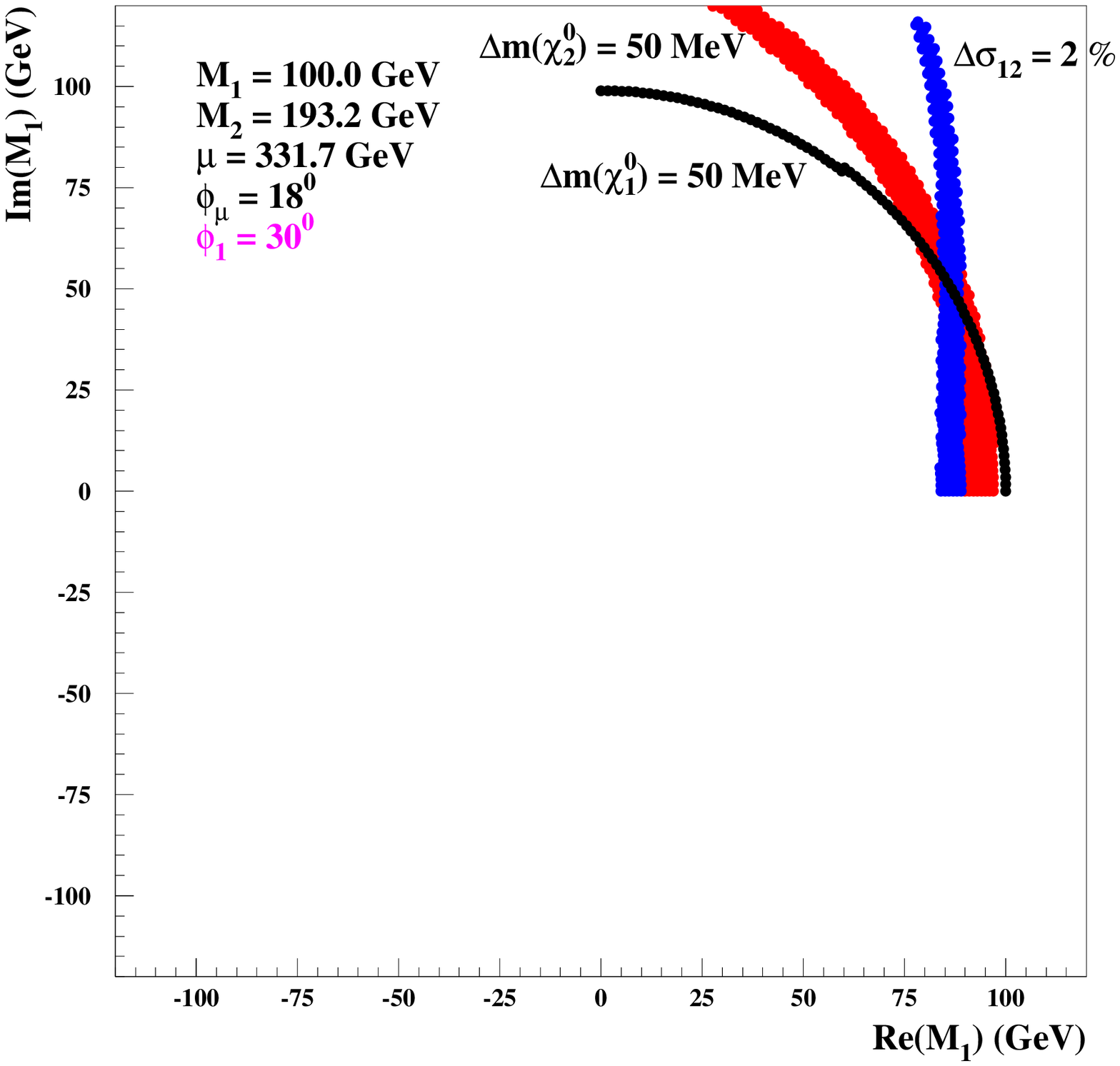}}
\end{picture}
\end{center}
\end{minipage}\vspace*{-.8cm}
\begin{minipage}{7cm}
\begin{center}
\begin{picture}(15,6)
\put(0,0){\includegraphics{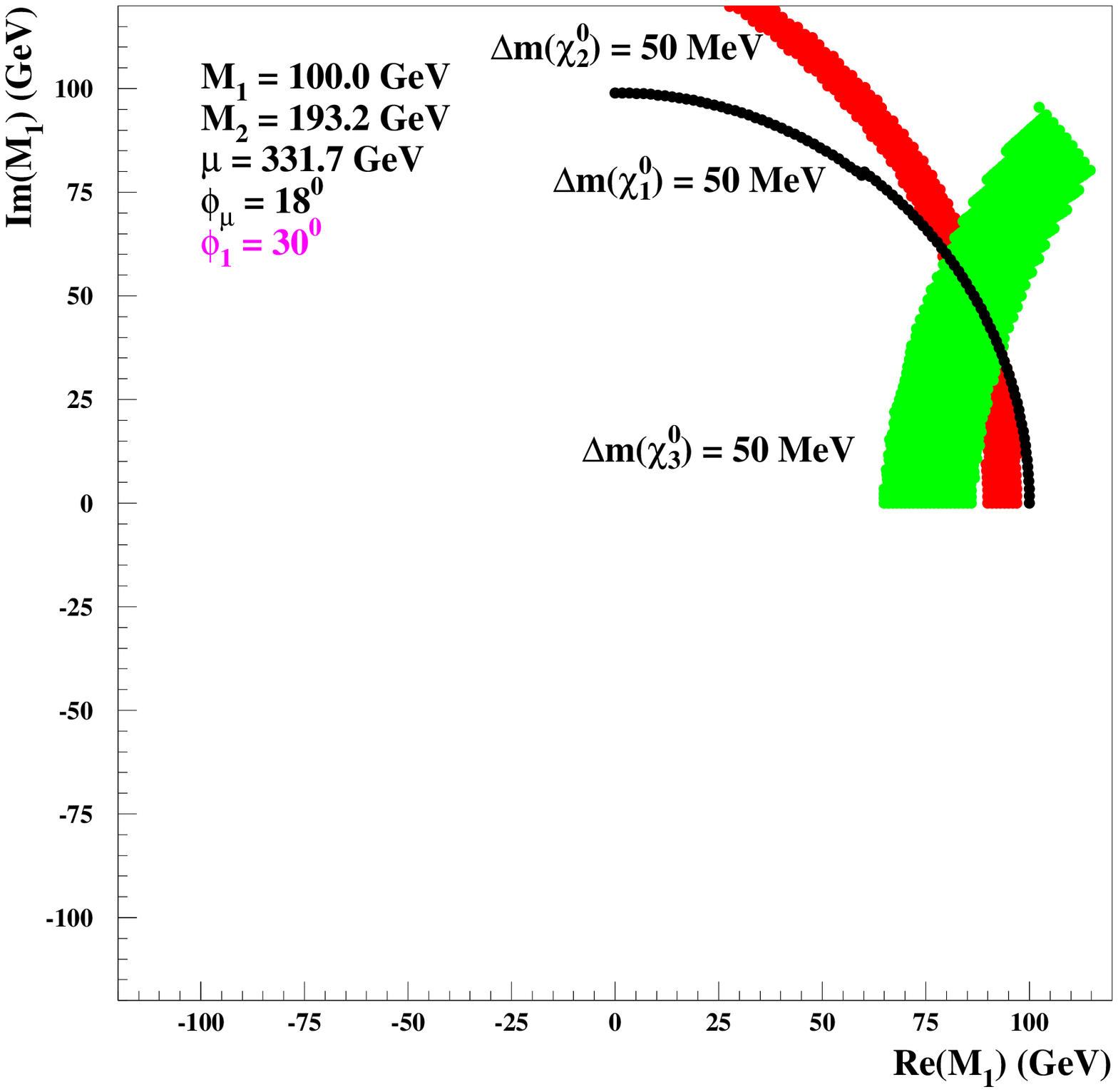}}
\end{picture}
\end{center}
\end{minipage}
\vspace*{-.3cm}
  \caption{\label{fig_2} Contour lines in the $Re(M_1)-Im(M_1)$ plane for 
the case where a) the two lightest masses $m_{\tilde{\chi}^0_{1,2}}$  and 
the cross section $\sigma(e^+e^-\to\tilde{\chi}^0_1\tilde{\chi}^0_2)$ are 
measured (left) and
b) the three lightest masses $m_{\tilde{\chi}^0_{1,2,3}}$   
are measured (right). It has been 
assumed $\delta(m_{\tilde{\chi}^0_i})\sim 50$~MeV and the 
statistical uncertainty for $\sigma$ \ci{Desch}.
  }%
\end{figure}

For an unambiguous determination (up to a simple
sign ambiguity in the phase) 
one therefore needs either three neutralino
masses or two masses and one cross section to resolve the
ambiguity. We investigate which of these possibilities would lead
to a higher accuracy for the determination of $M_1$. We
compare the two cases, taking into account the expected errors for
neutralino mass measurements at TESLA done for a given SUSY scenario
\cite{TDR} and the statistical error
of the measured cross sections \ci{Desch}, see Figs.~\ref{fig_2}a,~b.
The experimental constraints for
$\Phi_{M_1}$ are weaker than those for $\Phi_{\mu}$,  
so that a relatively large phase for
$M_1$ can not be excluded a priori. This is different for $\Phi_{\mu}$, where
the experimental constraints
for the dipole moments of the electron, neutron and mercury
atom are rather strict (see \cite{Abel} and references therein). 

We see from Figs.~\ref{fig_2}a,~b that one can determine the 
phase $\Phi_{M_1}$ about one order of magnitude more accurate 
when studying the two light
masses and the corresponding polarized cross section,
$\Phi_{M_1}=30^{0}\pm 3^0$, as compared to the case when three masses are 
studied, $\Phi_{M_1}=30^{0}\pm 10^0$.

\subsection{Parameters from only light Charginos/Neutralinos}
\begin{figure}
\setlength{\unitlength}{1cm}
\begin{center}
\begin{picture}(15,10)
\put(0,0){\includegraphics{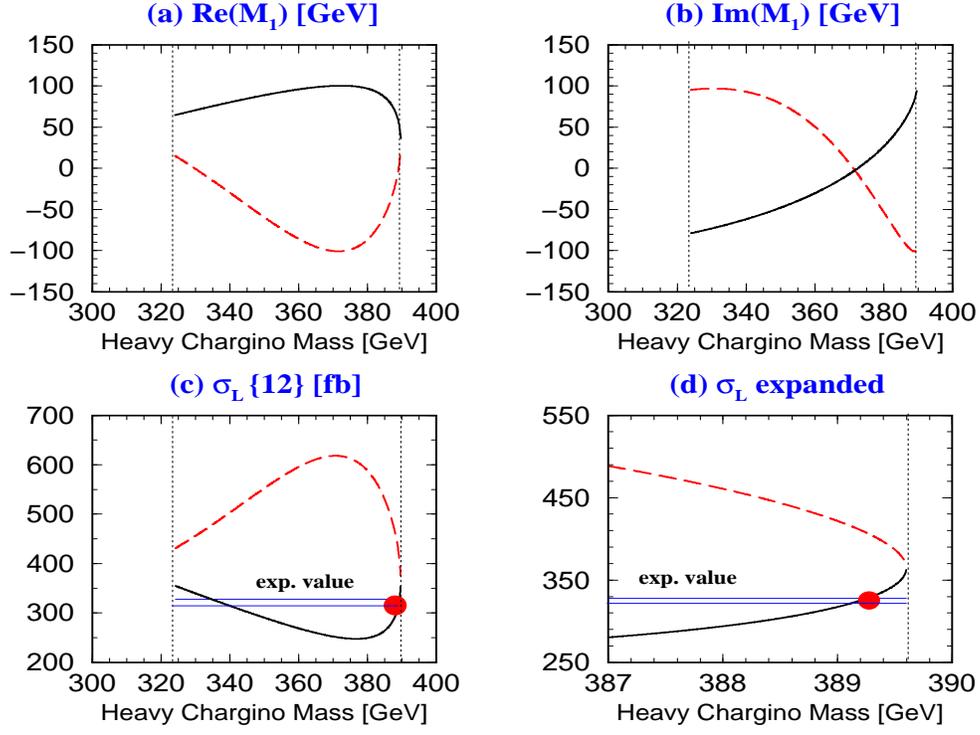}}
\end{picture}
\end{center}\vspace*{-2cm}
  \caption{\label{fig_3} The parameter set ($Re(M_1)$, $Im(M_1)$) 
and the prediction 
for the cross sections of neutralino production 
$\sigma_L(e^+e^-\to\tilde{\chi}^0_1\tilde{\chi}^0_2)$ 
as function of 
$m_{\tilde{\chi}^{\pm}_2}$ for the crossing points of the two circles in 
Fig.~\ref{fig_1}.}%
\end{figure}

In the case where one could only measure $\tilde{\chi}^{\pm}_1$ and the
polarized cross sections $\sigma_{L,R}(e^+e^-\to \tilde{\chi}^+_1
\tilde{\chi}^-_1)$ it is not possible to determine the parameters
$M_2$, $\mu$, $\phi_{\mu}$ and $\tan\beta$ uniquely. Instead of the two
crossing points, Fig.~\ref{fig_1}, one gets two parameter samples
as function of the heavier unknown mass
$m_{\tilde{\chi}^{\pm}_2}$. Since charginos are a $2\times 2$ system
one can set bounds for $m_{\tilde{\chi}^{\pm}_2}$:
\bequ
\frac{1}{2}\sqrt{s}-m_{\tilde{\chi}_1^{\pm}}\le
m_{\tilde{\chi}^{\pm}_2}\le\sqrt{m_{\tilde{\chi}_1^{\pm}}^2
+4 m^2_W/|\cos 2 \Phi_L-\cos 2 \Phi_R|}. 
\label{mixing}
\eequ
We show in Fig.~\ref{fig_3}a,~b the 
parameters $Re(M_1)$ and $Im(M_1)$ as 
function of $m_{\tilde{\chi}^{\pm}_2}$. 
In order to fix the parameters one has to explore in addition
polarized cross sections for neutralino production
$\sigma_{L,R}(e^+e^-\to\tilde{\chi}^0_1\tilde{\chi}^0_2)$, 
Fig.~\ref{fig_3}c,~d. With this procedure one gets in addition to the 
determination of the parameters $M_1$, $\Phi_{M_1}$
also a prediction for the
heavier mass $m_{\tilde{\chi}^{\pm}_2}$. This is done by
comparing  
the theoretical prediction for the cross sections
with the measured rates for $e^+e^-\to\tilde{\chi}^0_1\tilde{\chi}^0_2$ .

\begin{figure}
\setlength{\unitlength}{1cm}
\begin{center}
\begin{picture}(15,3)
\put(0,0){\includegraphics{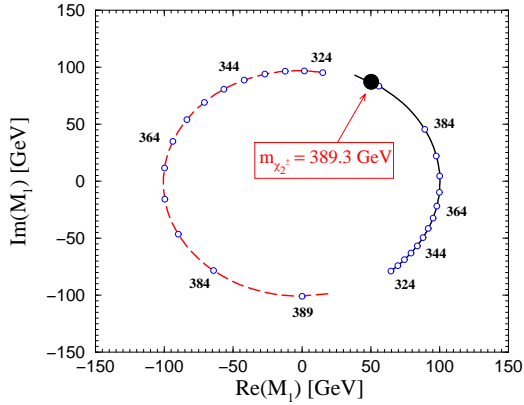}}
\end{picture}
\vspace*{-1.5cm}
  \caption{\label{fig_4} The trajectories of the two crossing 
points of $m_{\tilde{\chi}^0_1}$, $m_{\tilde{\chi}^0_2}$ as
function of $m_{\tilde{\chi}^{\pm}_2}$. The thick dotted point
denotes the correct solution where the theoretical prediction for the cross
section coincides with its measured value.}%
\end{center}
\end{figure}

The procedure for the parameter determination from only the
light system is illustrated in Fig.~\ref{fig_4} where the 
trajectories of the two crossing points of $m_{\tilde{\chi}^0_1}$,
$m_{\tilde{\chi}^0_2}$ are given in the $Re(M_1)$, $Im(M_1)$ plane as
function of $m_{\tilde{\chi}^{\pm}_2}$.  The thick dotted point
denotes the correct solution where the theoretical prediction for the cross
section coincides with its measured value \cite{CKMZ}.

\vspace*{-.5cm}
\section{Conclusions and Outlook}
\vspace*{-.3cm}
A future Linear Collider will be well suited to discover and reveal
precisely the underlying structure of the MSSM.  We have shown how to
determine the fundamental parameters of the chargino and neutralino mixing 
matrices: the parameters $M_2$, $\mu$, $\Phi_{\mu}$ and moderate
$\tan\beta$ can be determined via the chargino sector. The parameters
$M_1$, $\Phi_{M_1}$ can be determined by measuring two neutralino
masses and one cross section with high precision at a LC. The proposed
procedure works also in the case where only the lightest chargino and
the two lightest neutralino are accessible
and leads in addition to 
a prediction for $m_{\tilde{\chi}^{\pm}_2}$.

If $\tan\beta>10$, however, the chargino/neutralino
sector is rather insensitive to this parameter. The shown
procedure could then be explored 
in combination with the $\tau/\tilde{\tau}$ system leading
also in this case
to an accurate parameter determination (see also \cite{Boos}).

\vskip .5em
\begin{sloppypar}
The author would like to thank S.Y.~Choi, K.~Desch, 
J.~Kalinowski and P.M.~Zerwas for fruitful collaboration.
\end{sloppypar}

\end{document}